# Einstein and the Formal Equivalence of Mass and Energy

Paul A. Klevgard, Ph.D.
Sandia National Laboratory, Ret.
pklevgard@gmail.com

**Abstract**

This is a brief look at how Einstein explored formal symmetries between quantized matter and quantized radiation between 1903 and 1925. Specifically he employed thermodynamic comparisons between the ideal molecular gas and the photon gas. His achievements are tied in with a more general pattern in physics to explore formal symmetries between quantized mass and quantized energy.

There is a priceless cartoon of a grade school Einstein handing in an assignment with nothing on the paper but his name and $E = mc^2$. The second frame of the cartoon shows the paper after the teacher has graded it and her note at the bottom says: "Good effort Albert. Next time show your work, C minus."

Of course the grown up Einstein did show his work in his 1905 paper[1] but he made no comments there on how his equation was to be applied. Succeeding generations of physicists have interpreted it strictly in quantitative terms: a certain quantity of mass is convertible to a certain quantity of energy and vice versa. But Einstein himself found a deeper, more fundamental connection between mass and energy, something akin to a formal symmetry. His application of this was usually within the context of thermodynamics and the properties of the blackbody radiation gas versus the ideal molecular gas. Other physicists have found symmetries between mass and energy, between the particle and the photon, but Einstein's efforts in this regard deserve recounting here. Other writers have covered much of this in greater detail, but let this brief retelling serve to underscore the point that mass and energy have deeper connections than quantitative convertibility.

Einstein's earliest interests in physics centered on thermodynamics, statistical mechanics and kinetic theory. His creative research on these subjects spanned a quarter century and some 40 articles, and all of his principal contributions to quantum theory were statistical in origin.[2] In his famous paper of March 1905, Einstein introduced and applied to the photoelectric effect the hypothesis that light consists of discrete energy quanta. Einstein based his light quantum theory on arguments drawn from statistical mechanics, arguments whose novelty (that of symmetry) is reflected in the title of the paper, "On a Heuristic Viewpoint Concerning the Production and Transformation of Light." Commentators often characterize the arguments of the paper as creating an analogy between radiation and a classical ideal gas of material particles. But to Einstein, the connection was more than an analogy; it was a "far-reaching formal relationship."[3]

---

[1] A. Einstein, "Does the Inertia of a Body Depend Upon Its Energy Content?". Translation by George Barker Jeffery and Wilfrid Perrett in *The Principle of Relativity*, London: Methuen and Company, Ltd. (1923).
[2] Abraham Pais, 'Subtle is the Lord...'The Science and the Life of Albert Einstein, (New York, 1982), p. 56.

Einstein began with a derivation of the entropy decrease when $N$ discrete molecules of an ideal gas contract from a larger volume $v_1$ to a smaller volume, $v_2$. The probability expression for this contraction is $W = (v_2/v_1)^N$. Using Wien's radiation function relating energy density, temperature, and frequency, Einstein derived an expression for the entropy decrease when monochromatic radiation of energy $E(q)$ in volume $v_1$ is squeezed into a smaller volume $v_2$. He showed the probability expression for this latter process is $W = (v_2/v_1)^{(E(q)/hq)}$. Einstein then proposed that the parallelism of mathematical form and thermodynamic behavior between the two cases suggest that radiation is identical in form to an ideal gas, and that it is composed of discrete particles, namely, energy quanta. Specifically, the number of radiation energy quanta $N$ must be equal to $E(q)/hq$, hence, $E(q) = Nhq$. A system of $N$ quantum units then has $hq$ energy units associated with each "particle." This is a brilliant, inspired hypothesis that had to wait more than a decade for experimental confirmation.

In 1909, Einstein published two papers expanding on his view that light is composed of energy quanta. One of his arguments again drew on the thermodynamic relations connecting an ideal gas with radiation, this time from a black body. He imagined a flat plate as a perfect reflector buffeted simultaneously by the collisions from gas molecules and by the pressure from blackbody radiation. The argument joined together his earlier work on molecular Brownian motion with his central concept of light quanta as a type of radiation gas.

Two more papers followed in 1916-17, again focusing on blackbody radiation. Einstein began with a reference to the ideal gas, noting the formal resemblance between the frequency distribution of blackbody radiation and Maxwell's distribution law for the velocities of gas molecules. This observation was by way of justifying one of his favorite procedures, namely, considering a molecular gas in equilibrium with blackbody radiation. Such a thought experiment had molecules both emitting and absorbing quantized radiation energy while still maintaining energy equilibrium. Combining these circumstances, a few generalized assumptions about emission and absorption, and the statistical interpretation of the second law of thermodynamics, Einstein brought forth a new derivation of Planck's law. He then addressed the problem of how a Maxwell distribution of molecular velocity could maintain itself subject to the effects of radiation pressure. In other words, how would an ideal gas and a photon gas interact in the equilibrium state in reference to molecular velocity distribution? The answer led Einstein to postulate that light quanta carry a momentum of *hq/c*, a hypothesis anticipated by Stark (1909) and subsequently confirmed by Compton in 1923. As in 1905 and 1909, Einstein's papers in 1916-17 on radiation quanta again depended upon the formal relationships between the molecular and the radiation gases.

The 1917 paper was Einstein's last on radiation, although he never ceased to ponder the mysteries of light quanta.[4] By the early 1920s, he was deeply involved in his unified field theory, and he was not a direct participant in quantum theory research. But in the summer of 1924, Einstein received from an Indian physicist, S. N. Bose, a manuscript whose publication Einstein arranged. In this paper,

---

[3] A. Einstein, "Quantentheorie des einatomigen idealen Gases. 2. Abhandlung," *Sitzber. Ber. Preuss. Ak. Wiss.* (1925): 3.

[4] In a letter to Besso in 1951, Einstein wrote: "All these fifty years of pondering have not brought me any closer to answering the question, 'What are light quanta?'"

Bose elaborated a new technique of counting the statistical distribution of total energy E over N particles (light quanta). The result constituted "a natural development of the ideas that Einstein had been advocating for close to twenty years...."[5] For a few months, Einstein put aside his work on the unified field and turned back to statistical mechanics. But whereas Bose used his new counting procedure merely to derive Planck's law, Einstein used the procedure to deepen the relationship he knew existed between molecular and photon gases.

In three papers of 1924-25, Einstein treated the molecular gas, not the photon gas, as he moved into the uncharted waters of quantum statistics. He analyzed and explained the low temperature gas phenomenon now known as the Bose-Einstein condensation. This analysis was immensely satisfying to him on the grounds of symmetry: a Bose-Einstein molecular gas yielded the third law of thermodynamics, just as a Bose-Einstein photon gas yielded Planck's law. But the most important achievement of Einstein's last creative hurrah led him close to a topic about which he had never felt entirely comfortable, wave mechanics.

When exploring the energy fluctuations of the photon gas in 1909, Einstein discovered an expression with two terms, one dependent upon radiation's continuous (wave interference) nature, and the other, surprisingly, dependent upon radiation's presumed discrete nature. In 1924, his conjugal analysis examined the density fluctuations of the molecular gas. This time, he found the expected term deriving from discrete effects; but in addition, there appeared a term suggesting continuous/wave effects. Having read de Broglie's thesis some months earlier, Einstein did not hesitate in associating a wave field with the rapidly moving gas molecules. Einstein's original ideas of fluctuations and duality had now come full circle. The energy fluctuations in the photon gas were mirrored in the density fluctuations of the molecular gas; and the particle/wave character of photons had its counterpart in the wave/particle nature of gas molecules.

Einstein's acceptance of de Broglie's wave/particle duality for matter, and his use of it in his quantum theory of the ideal gas, broke the logjam that had been impeding quantum theory development. For some years, Schrödinger, too, had attempted to apply quantum theory to the ideal gas. He read Einstein's papers and saw a way to use de Broglie ideas to evade Bose's strange counting procedure. In an article entitled "On Einstein's Gas Theory," completed in late 1925, Schrödinger viewed the gas energy as blackbody radiation waves within a cavity. Whereas Einstein and others distributed the energy of the gas over a statistical array of indistinguishable molecules, Schrödinger regarded the gas as a collection of harmonic oscillators whose energy was distributed over various modes of oscillation. A month later, Schrödinger applied these same ideas of harmonic, resonant oscillation to the hydrogen atom, and wave mechanics was born.

Spanning two decades of research and publishing, Einstein's central idea was the relationship of the photon gas to the molecular/ideal gas. He was able to derive physical laws and testable predictions from the symmetrical thermodynamic relationships of the two gases, one of mass quanta and one of radiation quanta. Einstein recognized the differences of the two gases in content but believed they shared the same formal principles.

---

[5] Martin J. Klein, "Einstein and the Wave-Particle Duality," *Natural Philosopher* 3 (1964): 27.

Einstein was not the only physicist who thought that mass and energy had similarities at the foundational level. Both de Broglie and Schrödinger, inspired by Einstein's writings and examples, pushed the mass - energy symmetry even further. Together they showed that quantized particles had wave properties just as quantized radiation had particle properties. Preceding all of them was William Rowan Hamilton whose mechanics was based on an analogy between the classical mechanics of particles and the physics of light rays. Interpreting $E = mc^2$ as merely quantitative is to ignore a persistent trend in physics to find other meaningful relationships between quantized mass and quantized energy.